\documentclass[aip,amsmath,amssymb,reprint]{revtex4-2}
\usepackage[normalem]{ulem} % for strikethrough
\usepackage{xcolor}         % for colors

\usepackage{color}
\usepackage{hyperref}
\usepackage{enumitem}
\usepackage{graphicx,subfigure}
\usepackage[normalem]{ulem}
\usepackage{lineno}
\usepackage{booktabs}
\usepackage{balance}
\usepackage{array}
\usepackage{multirow}
\usepackage{hhline}
\usepackage{makecell}
\usepackage[symbol]{footmisc}
\usepackage{siunitx} % Required for alignment
\usepackage{comment}
\usepackage{float}
\usepackage{natbib}
\usepackage{multibib}
\newcites{SI}{References}

\usepackage{chemformula} % Formula subscripts using \ch{}
\usepackage[T1]{fontenc} % Use modern font encodings

\usepackage{soul} % for \ULon and \markoverwith
\newcommand{\redout}[1]{%
  \bgroup
  \markoverwith{\textcolor{red}{\rule[0.5ex]{2pt}{0.5pt}}}%
  \ULon{#1}%
  \egroup
}
\begin{document}
%%%%%%%%%%%%%%%%%%%%%%%%%%%%%%%%%%%%%%%%%%%%%%%%%%%%%%%%%%%%%%%%%%%%%
%% Meta-data block
%% ---------------
%% Each author should be given as a separate \author command.
%%
%% Corresponding authors should have an e-mail given after the author
%% name as an \email command. Phone and fax numbers can be given
%% using \phone and \fax, respectively; this information is optional.
%%
%% The affiliation of authors is given after the authors; each
%% \affiliation command applies to all preceding authors not already
%% assigned an affiliation.
%%
%% The affiliation takes an option argument for the short name.  This
%% will typically be something like "University of Somewhere".
%%
%% The \altaffiliation macro should be used for new address, etc.
%% On the other hand, \alsoaffiliation is used on a per author basis
%% when authors are associated with multiple institutions.
%%%%%%%%%%%%%%%%%%%%%%%%%%%%%%%%%%%%%%%%%%%%%%%%%%%%%%%%%%%%%%%%%%%%%
\author{Yuvraj Singh}
\affiliation{Department of Physics, Indian Institute of Science Education and Research (IISER) Tirupati, Tirupati, Andhra Pradesh, 517619, India}
\author{Chandan K. Choudhury}
\affiliation{Prescience Insilico Pvt. Ltd. Bengaluru, Karnataka, 530068, India }
\author{Rakesh S. Singh}
\affiliation{Department of Chemistry, Indian Institute of Science Education and Research (IISER) Tirupati, Tirupati, Andhra Pradesh, 517619, India}
\email{rssingh@iisertirupati.ac.in}

%%%%%%%%%%%%%%%%%%%%%%%%%%%%%%%%%%%%%%%%%%%%%%%%%%%%%%%%%%%%%%%%%%%%%
%% The document title should be given as usual. Some journals require
%% a running title from the author: this should be supplied as an
%% optional argument to \title.
%%%%%%%%%%%%%%%%%%%%%%%%%%%%%%%%%%%%%%%%%%%%%%%%%%%%%%%%%%%%%%%%%%%%%
\title
  {From Fresh to Salty: How Ions Modulate Solvent-Mediated Interactions between Grafted Silica Nanoparticles in Water}

\begin{abstract}
Nanoparticles (NPs) are fundamental building blocks for engineering functional soft materials, where precise control over the solvent-mediated inter-particle effective interaction ($U_{\rm eff}$) is essential for tailoring bulk structure and properties. These solvent-mediated interactions are strongly influenced by NP’s surface chemistry, solvent properties, and thermodynamic conditions such as temperature ($T$) and pressure ($P$). However, despite considerable progress, a general predictive framework for tuning $U_{\rm eff}$ and guiding self-assembly remains lacking. In this work, using all-atom molecular dynamics simulations, we investigated the alteration of $U_{\rm eff}$ between silica nanoparticles (Si-NPs) functionalized with polyethylene (PE) and polyethylene glycol (PEG) by salt (sodium chloride) across a range of thermodynamic conditions. At ambient thermodynamic conditions, bare (not functionalized) Si-NPs exhibit minimal variation in $U_{\rm eff}$ even at high salt concentrations (up to $5$ molal). In contrast, PE-grafted Si-NPs display strong salt-induced attractions, while PEG-grafted Si-NPs show an intermediate, more gradual response. To asses the transferability of these salt-induced effects on effective interactions, we further examined the effects of salt on $U_{\rm eff}$ under different ($T,P$) conditions. Our results indicate that the salt-induced modulation of $U_{\rm eff}$ between both bare and grafted Si-NPs is largely invariant across the explored ($T,P$) conditions. Molecular-level analysis reveals that salt promotes solvent depletion within the interparticle cavity for both hydrophobic PE and hydrophilic PEG grafts, with the strongest effect observed in the PE case. In general, this study highlights the coupled roles of surface chemistry, ion-polymer interactions, and solvent structuring in the regulation of $U_{\rm eff}$, and provides important insights into the predictable control of interparticle interactions for soft material engineering. 
\end{abstract}
\maketitle

\section{Introduction}
Nanoparticle (NP) self-assembly has emerged as a promising approach in materials science that allows bottom-up fabrication of complex materials with tailored properties~\cite{glotzer2007anisotropy, fan2010self, grzelczak2010directed, edel2013self, zhou2016biomimetic}. This methodology is central to a wide range of applications such as plasmonics~\cite{fan2010self, willets2007localized}, photovoltaics~\cite{karg2015colloidal}, chemical and biological sensing~\cite{saha2012gold, karnwal2024gold}, and drug delivery~\cite{yang2021designing, mitchell2021engineering}. In these systems, NPs act as fundamental building blocks where controlling their self-assembly pathways is essential to achieve targeted structural and functional outcomes. There is a growing interest in developing inverse design methods with the aim of achieving specific interaction potentials that guide the self-assembly process to the target structure. Self-assembly pathways can be systematically modulated by tuning interparticle interactions by altering intrinsic NP characteristics such as size, shape, and surface chemistry or by the surrounding solvent environment~\cite{chen2012janus, liepold2019pair, singh2024computational, hummer1998pressure, li2017strong}. Such tunability of local interparticle interactions is central to rational design of self-assembled materials~\cite{lu2012colloidal,linko2022precision, baumann2019metal, bianchi2017limiting}. 

Computer simulations play a crucial role in this effort by providing detailed molecular-level insights into how nanoscale building blocks (such as NPs or colloids) dispersed in a solvent interact and organize. In many systems, the solvent-mediated effective interaction ($U_{\rm eff}$) between NPs is not solely determined by their intrinsic properties (such as size, shape, and surface chemistry) but is strongly influenced by the surrounding solvent~\cite{galteland2020solvent, bresme2009solvent, prince2020solvent,thirumalai2012role,kopel2019comparative}. By tuning the solvent-mediated interactions --- either through NP's surface modifications or by altering solvent properties --- one can effectively control self-assembly pathways and engineer structures with tailored properties. The surface characteristics of NPs can be modulated by adjusting surface polarity, for example via polymer or DNA grafting~\cite{peng2013engineering, subjakova2021polymer, ma2022dna, parolini2016controlling, sciortino2020combinatorial,yu2023molecular}, whereas the solvent properties can be tailored by introducing additives such as salts~\cite{benavides2016consensus, ali2020salt, ding2014anomalous, blazquez2023computation} or alcohols~\cite{ghosh2016temperature, noskov2005molecular, zhang2006mutual}. In the case of grafted NPs, the interplay between the solvent and the grafted polymers, as well as, solvent-NP interactions play a crucial role. For example, favorable solvent-polymer interactions typically lead to extended polymer conformations, while unfavorable interactions can induce polymer collapse or adsorption onto the NP surface~\cite{linse2012effect, polanowski2025monte, mukherji2014polymer} resulting in a change in inter-NP effective interactions. These effects are not only substantial but also highly complex and depend on the system under consideration. Furthermore, thermodynamic conditions such as pressure ($P$) and temperature ($T$) can significantly influence the behavior of both solvent and polymer, thereby altering $U_{\rm eff}$ between NPs~\cite{salas2024pressure, yadav2016effective, hatakeyama1974compressibility, herrera2025effect}. Variations in temperature can change the solvent's behavior, affecting whether it promotes polymer swelling or collapse~\cite{backes2017poly, piguet2016high, mrad2023polyethylene,alizadeh2021systematic}. Similarly, pressure can influence molecular packing and solvent density, which in turn modifies solvent properties and polymer conformations~\cite{de2015does, tsujita1973thermodynamic, capt2000pressure, fontana2010high}. These changes directly impact how NPs interact and aggregate in the medium, and therefore, understanding these effects requires elaborate computational efforts to decipher the interplay between solvent properties, NP's surface characteristics and effective inter-particle interactions.  

Recently, considerable effort has been devoted to understanding how the free energy landscape (FEL) governs self-assembly and phase transition pathways~\cite{wolde1997enhancement, ramesh2024microscopic, iwamatsu2011free, santra2013nucleation, santra2018polymorph, singh2025manifestations, singh2023anomalous}. However, the role of effective pair interactions between constituent particles in shaping this landscape --- particularly in the context of competing multiple phases --- remains less explored. Recent simulation studies have shown that careful tuning of interparticle interactions can lead to a wide range of complex and hierarchical structures~\cite{curatolo2023computational, jacobs2015rational, grunwald2014patterns, bohlin2023designing, jain2014inverse, miskin2016designing, han2019reinforcement} which are local minima in the FEL. For example, Truskett and coworkers ~\cite{lindquist2016communication} introduced a simulation method to adjust pair interactions, promoting the formation of specific two-dimensional lattices. This approach was later extended to three-dimensional structures using a relative entropy optimization framework~\cite{lindquist2018inverse}. Similarly, Yu and Guo~\cite{yu2023molecular} demonstrated that the self-assembly of polymer-grafted nanocrystals is governed by a competition between the packing entropy of the grafted polymers and the directional entropy of the core shape, enabling programmable transitions between face-centered cubic (FCC), body-cenetrated cubic (BCC), and anisotropic superlattices. 

There is a growing experimental interest in understanding how NP surface functionalization influences their assembly into ordered structures. For instance, Rogers et al.~\cite{rogers2016using} demonstrated that DNA-grafted colloids can be programmed to form well-defined crystal lattices through sequence-specific interactions. Akcora et al.~\cite{akcora2009anisotropic} showed that polymer-grafted NPs can self-assemble into anisotropic superlattices by tuning polymer chain length and grafting density. In another example, Wang et al.~\cite{wang2019nanoimprint} reported that replacing a gold core with TiO$_2$ in polystyrene-grafted NPs leads to spontaneous clustering driven by enhanced interparticle attractions. To gain molecular-level insight into how surface functionalization modulates interparticle interactions and, in turn, self-assembly pathways, we recently performed simulations to examine the effect of short polymer grafts on the effective interaction $U_{\rm eff}$ between silica NPs (Si-NPs) dispersed in water~\cite{singh2024computational}. Using polymers spanning hydrophobic to hydrophilic character --- polyethylene (PE), polyethylene glycol (PEG), and polymethyl methacrylate (PMMA) --- we found that both the strength and range of $U_{\rm eff}$ are highly sensitive to polymer chemistry and surface grafting density. These findings underscore the central role of surface functionalization in reliably tuning interparticle forces. Despite these advances, the influence of external additives (such as salts) on solvent-mediated NP interactions remains comparatively underexplored, particularly the molecular mechanisms by which such additives modify $U_{\rm eff}$.

In the present study, we employed all-atom molecular dynamics (MD) simulations to systematically investigate the influence of sodium chloride (NaCl) salt under different ($T,P$) conditions on $U_{\rm eff}$ between bare (or, ungrafted) and grafted Si-NPs. Our results show that salt addition significantly alters the (effective) interaction strength, with the extent of this alteration depending on nature of surface functionalization (\textit{i.e.}, hydrophilic vs hydrophobic grafting polymers). The salt-induced effects are more pronounced for hydrophobic PE-grafted Si-NPs. We further investigated the transferability of these salt-induced effects by examining the effects of salt on $U_{\rm eff}$ under different ($T,P$) conditions and surface functionalizations. To elucidate the underlying mechanisms, we also analyzed the roles of polymer-solvent interaction, ion affinity, and local solvent structuring in governing the observed changes in $U_{\rm eff}$. Overall, this study highlights the coupled roles of surface chemistry, ion-surface (including grafted polymer) interactions, and solvent in governing $U_{\rm eff}$, offering key insights for the rational and predictable tuning of interparticle interactions.
\section{Computational Method details}
\subsection{Model and MD simulation details}
All-atom MD simulations were performed using GROMACS~\cite{gromacs} (version 2021.5), coupled with PLUMED~\cite{plumed} (version 2.8.1). We first modeled bare Si-NP of diameter of $2$ nm using CHARMM-GUI~\cite{lopes2006development} and two polymer types, PE and PEG, each comprising five repeating units, were grafted onto the bare Si-NP surface to model the grafted Si-NPs. A total of $22$ polymer chains were randomly attached on each particle, that is, the surface grafting density of $1.75$ chains/nm$^2$. The complete modeling protocol and force field parameters are described in our previous work~\cite{singh2024computational}. Two such NPs were placed in a cubic simulation box of $12$ nm side length. This box was then solvated with TIP3P~\cite{TIP3P} water containing NaCl at concentrations of $2$ molal (m) and $5$ m (\textit{i.e.}, $2$ and $5$ moles of salt per kg of water). Force field parameters for NaCl compatible with TIP3P were adopted from Ref.~\emph{\citenum{yagasaki2020lennard}} that is validated against experimental salt solubility. All simulations were conducted under isothermal–isobaric ($NPT$) conditions at temperatures $300$ K and $350$ K, and at pressures of $1$ bar and $1000$ bar. Temperature was controlled using the Nosé–Hoover thermostat~\cite{nose} with a relaxation time of $0.4$ ps, and pressure was maintained using the Parrinello–Rahman barostat~\cite{parrinello1981polymorphic} with a relaxation time of $2$ ps. A cutoff of $1.2$ nm was applied for both van der Waals and short-range Coulomb interactions. Long-range electrostatics were treated using the particle mesh Ewald (PME) method~\cite{PME}. The rigid body constraints were implemented using the linear constraint solver (LINCS) algorithm~\cite{lincs}. Periodic boundary conditions were applied in all three spatial directions. The equations of motion were integrated using the Verlet algorithm with $2$ fs timestep.
\subsection{Potential of mean force (PMF) computation}
The $U_{\rm eff}$ between NPs in water is estimated by computing the potential of mean force (PMF) as a function of the center-of-mass distance ($r_{\rm com}$) between Si-NPs using the umbrella sampling method. The reaction coordinate was divided into $35$ overlapping windows, spaced by $0.12$ nm. A harmonic bias potential with a force constant of $1500$  kJ/mol nm$^2$ was applied to each window to ensure adequate histogram overlap. For configurations with $r_{\rm com} < 4$ nm, the systems were equilibrated for at least $200$ ns, and for larger separations, equilibration times ranged from $100-150$ ns depending on the thermodynamic condition and the nature of the grafting polymers. The initial $50\%$ of the sampling data from each window were discarded and the remaining data were divided into five equal segments. The weighted histogram analysis method (WHAM) was applied to each segment to construct five independent PMF profiles, which were then used to assess convergence and estimate statistical errors. The final PMF profile used for analysis corresponds to the average of these five profiles. We also examined the effects of system size and equilibration time on $U_{\rm eff}$ by reducing the system size (number of water molecules) from $N = 50000$ to $N = 32000$ molecules (corresponding to average box lengths of approximately $12$ nm and $10$~nm, respectively, under near-ambient thermodynamic conditions) and extending the system equilibration time to $400$~ns. The results exhibited the same qualitative trend, with $U_{\rm eff}$ changing approximately by maximum $10$\% near the contact separation for the PE-grafted case. 
\begin{figure*}
    \centering
    \includegraphics[width=\linewidth]{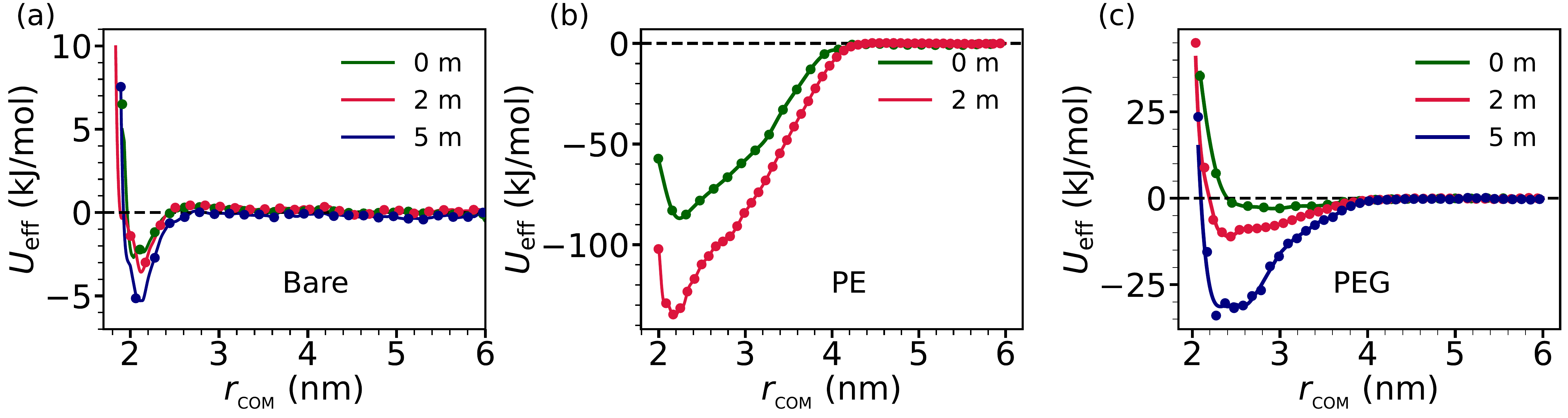}
    \caption{The salt-dependent behavior of $U_{\rm eff}$ as a function of $r_{\rm com}$ for bare (a), PE-grafted (b), and PEG-grafted (c) systems at $1$ bar and $300$ K is shown. For bare Si-NP case, the absolute change in the depth of $U_{\rm eff}$ with increasing salt concentration is minimal with an increase of $ \sim 3$ kJ/mol at $5$ m concentration of NaCl. In contrast, the PE-grafted system exhibits a pronounced increase in the depth of $U_{\rm eff}$. For the PEG-grafted case, the depth of the attractive basin increases to an intermediate extent compared to the bare and PE-grafted cases. Additionally, the hard repulsive core (a measure of the effective Si-NP diameter) shifts towards smaller $r_{\rm com}$ value.}
    \label{fig:1}
\end{figure*}
\section{Results and discussion}
\subsection{Salt-induced modulation of solvent-mediated effective inter-Si-NP interaction}
The solvent-mediated effective interaction $U_{\rm eff}$ between NPs --- characterized by the PMF --- describes the free energy profile as a function of the center-of-mass distance between two Si-NPs. We first computed $U_{\rm eff}$ to investigate how variations in salt concentration influence the interactions between Si-NPs in aqueous NaCl solution. Figure~\ref{fig:1} presents the $U_{\rm eff}$ between two Si-NPs as a function of $r_{\rm com}$ under different salt concentrations for bare, PE-grafted, and PEG-grafted systems at $300$ K and $1$ bar. For the bare Si-NP case (Fig.~\ref{fig:1}a), $U_{\rm eff}$ exhibits a shallow minimum near $r_{\rm com} \approx 2.1$ nm. The depth of this minimum (a measure of the strength of interaction) increases slightly from approximately $-2.5$ kJ/mol in pure water to  $-5.0$ kJ/mol as the salt concentration increases to $5$ m. This modest change suggests that the interaction between bare Si-NPs is only weakly sensitive to the presence of the NaCl salt. However, in contrast, the PE-grafted system (Fig.~\ref{fig:1}b) shows a much stronger response to the change in salt concentration. At $2$ m NaCl concentration, the depth of $U_{\rm eff}$ increases significantly to nearly $-127$ kJ/mol, compared to approximately $-82$ kJ/mol in salt-free solvent (pure water). Due to this substantial increase in attraction observed at the salt concentration $2$ m, no further simulations were performed at higher salt concentrations for this (PE-grafted) system. For PEG-grafted Si-NP system (Fig.~\ref{fig:1}c), the change in $U_{\rm eff}$ lies in between the bare and PE-grafted cases. The depth of $U_{\rm eff}$ increases from approximately $-2.5$ kJ/mol in a salt-free solvent to approximately $-25$ kJ/mol at a salt concentration of $5$ m. In addition, the repulsive core shifts towards a smaller diameter $r_{\rm com}$ with increasing salt concentration, suggesting a salt-dependent conformation change in grafted PEG polymers, resulting in a smaller effective diameter of the (grafted) Si-NP.  
\begin{figure}
    \centering
    \includegraphics[width=\linewidth]{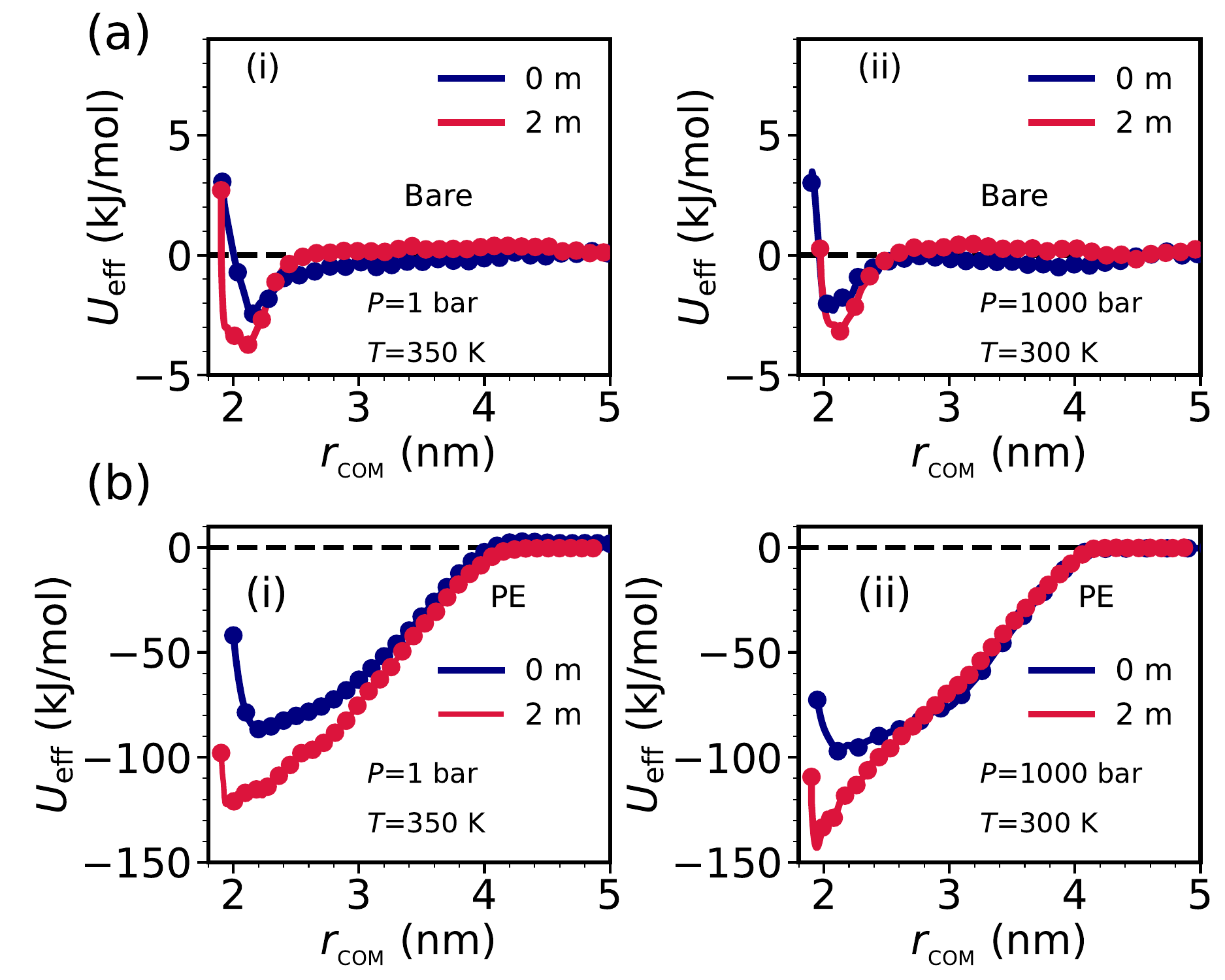}
\caption{The effects of salt on $U_{\rm eff}$ profile at elevated (with respect to ambient) pressure and temperature conditions is shown for the bare Si-NP (a) and PE-grafted Si-NP (b) systems. Here we have reported the $U_{\rm eff}$ profile at two thermodynamic conditions: $T = 350$ K, $P = 1$ bar and $T = 300$ K, $P = 1000$ bar; shown in (i) and (ii), respectively.  For the bare Si-NP system, $U_{\rm eff}$ remains largely unaffected on salt addition by increase of temperature from $300$ K to $350$ K and pressure to $1000$ bar. In contrast, the PE-grafted system shows a marked increase in the strength of attractive interaction on salt addition at elevated conditions as well.}
    \label{fig2}
\end{figure}
\begin{figure*}
    \centering
    \includegraphics[width=0.8\linewidth]{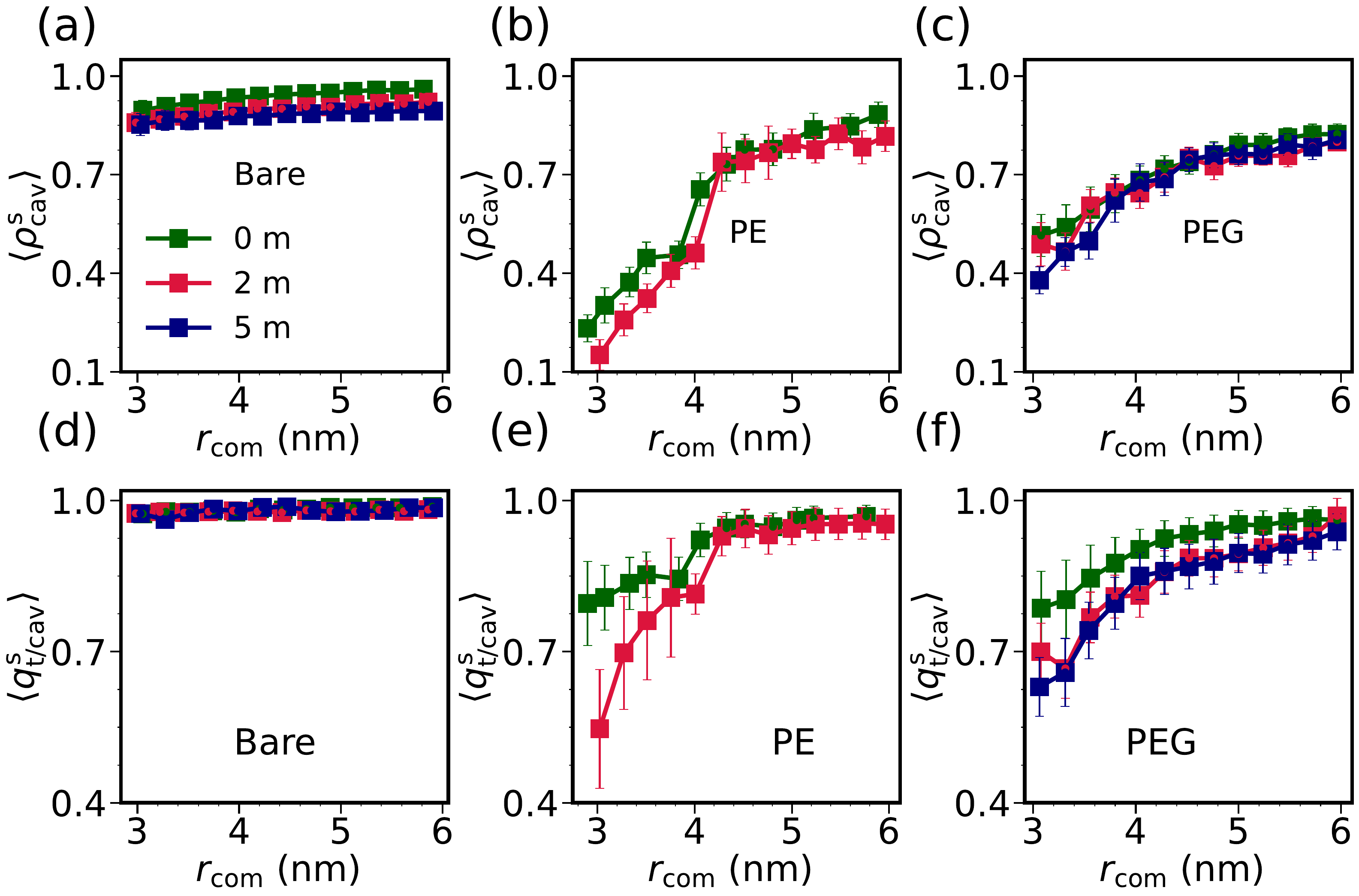}
\caption{Scaled average solvent number density $\langle \rho_{\rm cav}^{\rm s} \rangle$ (top, a-c) and scaled average tetrahedral order parameter $\langle q_{\rm t/cav}^{\rm s} \rangle$ (bottom, d-f) within the icavity as functions of inter-Si-NP center-of-mass separation $r_{\rm com}$ for bare, PE-grafted, and PEG-grafted Si-NPs at various salt concentrations. For PEG-grafted systems, $\langle \rho_{\rm cav}^{\rm s} \rangle$ decreases monotonically with increasing salt concentration, indicating enhanced solvent depletion. In the PEG-grafted case, a sharp decrease near $r_{\rm com} = 4$ nm at $2$ m salt suggests partial cavitation. The tetrahedral order parameter $\langle q_{\rm t/cav}^{\rm s} \rangle$ follows a similar trend to $\langle \rho_{\rm cav}^{\rm s} \rangle$, with water in the bare system largely retaining bulk-like structure.}
    \label{fig5}
\end{figure*}
\subsection{Is the salt-induced modulation of $U_{\rm eff}$ transferable across ($T,P$) conditions?}
Temperature and pressure are known to significantly influence effective interactions between nanoscale particles~\cite{FIEDLER20071786,Ferdous2012,salas2024pressure,Schroer2016} and surfaces~\cite{zangi2008temperature, ghosh2001molecular,PhysRevB.101.165432,PhysRevE.73.041604,Engstler2018,Engstler_jcp}. It is therefore important to examine whether the salt-induced modulation of $U_{\rm eff}$ observed near ambient conditions remains transferable across different thermodynamic ($T,P$) states --- that is, whether the extent of influence of salt on $U_{\rm eff}$ persists upon varying ($T,P$). To address this, we computed $U_{\rm eff}$ between two Si-NPs dispersed in solution under elevated ($T,P$) conditions. Figure~\ref{fig2} shows the modulation of $U_{\rm eff}$ by salt at $T = 350$ K and $P = 1$ bar, and at $T = 300$ K and $P = 1000$ bar for both bare, PEG- and PE-grafted Si-NPs. For bare Si-NPs, the $U_{\rm eff}$ profile exhibits only a modest change upon salt addition at elevated ($T,P$) conditions (see Figs.~\ref{fig2}a(i) and~\ref{fig2}a(ii)), consistent with the behavior observed under ambient conditions (Fig.~\ref{fig:1}(a)). This suggests that, in the low-salt regime, solvent-mediated interactions between bare Si-NPs are largely insensitive to variations in ($T,P$) within the ($T,P$) range probed in this work. The hydrophilic PEG-grafted Si-NPs also show a similar weak dependence on pressure (see Fig.~S1 in the Supplementary Material). 

In contrast, PE-grafted Si-NPs exhibit a much stronger response to salt under elevated ($T,P$) conditions, consistent with their behavior at ambient conditions. At $T = 350$ K and $P = 1$ bar, the depth of $U_{\rm eff}$ increases by approximately $42$ kJ/mol upon the addition of $2$ m salt (Fig.~\ref{fig2}b(i)), comparable to the enhancement observed under ambient conditions ($\sim 45$ kJ/mol; see Fig.~\ref{fig:1}b). At elevated pressure ($P = 1000$ bar, $T = 300$ K), we observe a slight increase of about $6$ kJ/mol in the salt-induced deepening of $U_{\rm eff}$ compared to the ambient case (from $\sim45$ kJ/mol near ambient pressure to $\sim51$ kJ/mol at 1000 bar). These results indicate that the salt-induced enhancement of effective attraction is largely transferable across thermodynamic conditions; in other words, variations in ($T,P$) exert only a weak influence on the salt-induced modification of $U_{\rm eff}$.

The results presented above (Sections~IIIA and~IIIB) highlight the intricate coupling between the nature of the grafting polymer (\textit{i.e.}, NP surface functionalization) and the ionic strength of the solvent in governing the effective pair interactions between nanoscale Si-NPs. These findings underscore the importance of considering both surface chemistry and solvent properties in tandem when assessing solvent-mediated forces between nanoscale building blocks.
\begin{figure*}
    \centering
    \includegraphics[width=0.82\linewidth]{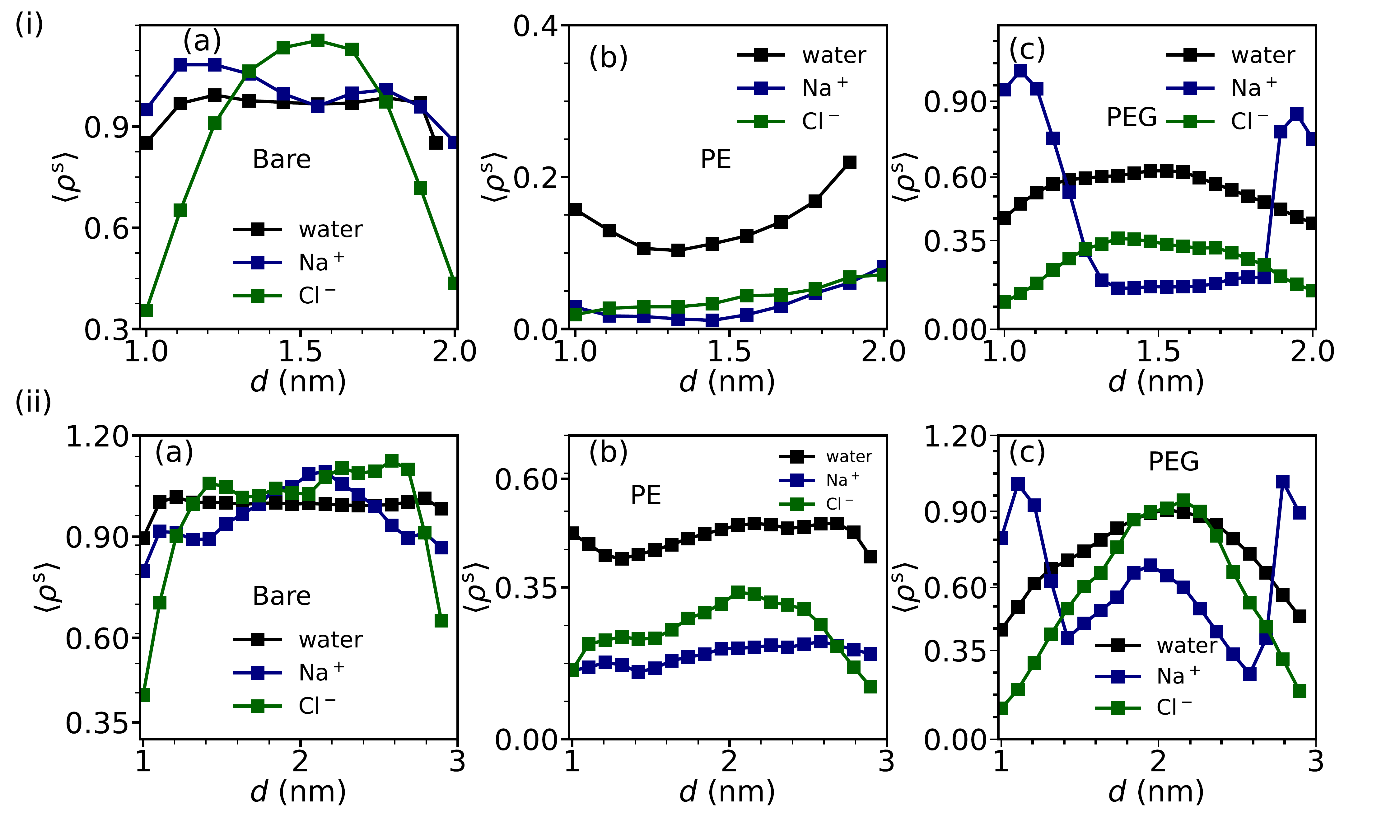}
    \caption{Average scaled ion and water number density ($\langle \rho^{\rm s} \rangle$) profile as a function of distance $d$ from the centre of a Si-NP along the axis connecting the centre-of-mass of two Si-NPs (see Fig.~S2 in the Supplementary Material) for $r_{\rm com} = 3.0$ nm (i) and $4.0$ nm (ii) and at $2$ m salt concentration. Here, we have reported the results for bare (a), PE-grafted (b), and PEG-grafted (c) Si-NPs at temperature $300$ K and pressure $1$ bar. The bare and PEG-grafted Si-NP systems show enhanced Na$^+$ density in the vicinity of the NP --- more pronounced for the PEG-grafted Si-NP indicating preferential Na$^+$ attraction with the PEG chains. The PE-grafted system exhibits relative lesser ion presence inside the cavity region for $r_{\rm com} = 4.0$ nm and negligibly small for $r_{\rm com} = 3.0$. Also, we report the absence of any preferential ion adsorption on the Si-NP surface for this case.}
    \label{fig4}
\end{figure*}
\section{Molecular origin of the alteration of the effective interactions between Si-NPs}
\subsection{Density and structural behavior of solvent confined between two Si-NPs}
The solvent-mediated interactions between particles strongly depend on the arrangement of solvent molecules around them and on how this arrangement evolves as the particles approach one another. To elucidate the molecular origin of the salt-dependent modulation of $U_{\rm eff}$ between Si-NPs (see, for example, Fig.~\ref{fig:1}), we analyzed the scaled average solvent number density ($\langle \rho_{\rm cav}^{\rm s} \rangle$), and the scaled average tetrahedral order of cavity water ($\langle q_{\rm t/cav}^{\rm s} \rangle$) of the solvent cavity formed between two Si-NPs (see Fig.~S2 in the Supplementary Material). The scaled average cavity number density is defined as, $\langle \rho_{\rm cav}^{\rm s} \rangle = \langle \rho^{\rm cav} \rangle / \langle \rho^{\rm sol} \rangle$, where $\langle \rho^{\rm cav} \rangle$ is the average number density of solvent (water + ions) inside the cavity and $\langle \rho^{\rm sol} \rangle$ is the corresponding bulk solvent density under the same ($T,P$) conditions and salt concentration. Similarly, the scaled tetrahedral order of cavity water is defined as, $\langle q_{\rm t/cav}^{\rm s} \rangle = \langle q_{\rm t}^{\rm cav} \rangle / \langle q_{\rm t}^{\rm bulk} \rangle$, where $\langle q_{\rm t}^{\rm cav} \rangle$ is the average tetrahedral order of water molecules in the cavity and $\langle q_{\rm t}^{\rm bulk} \rangle$ is the bulk value under identical thermodynamic and salt conditions. The local tetrahedral order parameter, $q_{\rm t}$, quantifies how closely the arrangement of a water molecule and its four nearest neighbors approaches an ideal tetrahedral geometry. For the $i^{\rm th}$ water molecule, it is given by~\cite{errington2001relationship},  
\begin{equation}
    q_{\rm t}(i) = 1 - \frac{3}{8} \sum_{j=1}^{3} \sum_{k=j+1}^{4} \left( \cos \psi_{jik} + \frac{1}{3} \right)^2,
\end{equation}
where $\psi_{jik}$ is the angle between the vectors connecting the central $i^{\rm th}$ water molecule to its $j^{\rm th}$ and $k^{\rm th}$ nearest neighbor. The $q_{\rm t} = 1$ corresponds to perfect tetrahedral coordination, while the lower values reflect a deviation from the perfect tetrahedral structure.

In Fig.~\ref{fig5} we show the salt-concentration dependence of $\langle \rho_{\rm cav}^{\rm s} \rangle$ and $\langle q_{\rm t/cav}^{\rm s} \rangle$ for bare, PE-grafted, and PEG-grafted Si-NPs at $1$ bar and $300$ K. We find that $\langle \rho_{\rm cav}^{\rm s} \rangle$ decreases systematically with increasing salt concentration, indicating progressive solvent depletion in the cavity region. This effect is most pronounced at shorter inter-particle separations, where confinement effects are stronger. For bare Si-NPs, $\langle \rho_{\rm cav}^{\rm s} \rangle$ decreases only modestly with salt concentration (Fig.~\ref{fig5}a), while $\langle q_{\rm t/cav}^{\rm s} \rangle$ remains nearly insensitive to salt concentration across all separations (Fig.~\ref{fig5}d). Thus, cavity water retains a near bulk-like tetrahedral structure at low salt concentrations, regardless of cavity size. This modest solvent depletion results in only a marginal change in the depth of $U_{\rm eff}$ ($\sim 2.8$ kJ/mol; Fig.~\ref{fig:1}a), corresponding to a weak enhancement in inter-particle attraction. By contrast, PE-grafted Si-NPs exhibit an abrupt decrease in $\langle \rho_{\rm cav}^{\rm s} \rangle$ near $r_{\rm com} \approx 4.5$ nm, consistent with a partial cavitation transition --- reminiscent of the cavitation or dewetting transition of water confined between two hydrophobic plates~\cite{cav_1, cav_2}. A similar pronounced decrease is also observed in $\langle q_{\rm t/cav}^{\rm s} \rangle$ at this separation, indicating a substantial disruption of the tetrahedral structure due to density loss. In our earlier study~\cite{singh2024computational}, we attributed the strong attraction between PE-grafted Si-NPs in pure water to such cavitation-induced interactions. Interestingly, the present results show that the cavitation transition becomes increasingly pronounced as the salt concentration rises from $0$ m to $2$ m, leading to a strongly enhanced cavitation-induced attraction ($\sim 45$ kJ/mol; see Fig.~\ref{fig:1}b).  

The PEG-grafted system displays a more intricate response to salt concentration. Due to its hydrophilic nature, PEG tends to retain water molecules in the cavity through favorable polymer--solvent interactions. As a result, in the salt-free case, the cavity solvent density $\langle \rho_{\rm cav}^{\rm s} \rangle$ remains relatively high even at close separations compared to the hydrophobic PE-grafted case, leading to only a marginal increase in the effective attraction~\cite{singh2024computational}. However, upon introducing salt, a sharper solvent depletion emerges near $r_{\rm com} \approx 4$ nm (Fig.~\ref{fig5}c), manifested in a distinct drop in both $\langle \rho_{\rm cav}^{\rm s} \rangle$ and $\langle q_{\rm t/cav}^{\rm s} \rangle$ (Figs.~\ref{fig5}c and~\ref{fig5}f). This depletion becomes more pronounced with increasing salt concentration and correlates directly with the enhancement of effective attraction (\textit{i.e.}, deepening of $U_{\rm eff}$ near contact) between PEG-grafted Si-NPs (Fig.~\ref{fig:1}c). Also, the shift of the repulsive core toward shorter $r_{\rm com}$ with increasing salt concentration can be attributed to enhanced cavitation, which facilitates a closer approach of the PEG-grafted Si-NPs. Thus, for all systems considered (bare and grafted Si-NPs), the separation-dependent behavior of $\langle \rho_{\rm cav}^{\rm s} \rangle$ and $\langle q_{\rm t/cav}^{\rm s} \rangle$ is strongly correlated with $U_{\rm eff}$, indicating that the origin of enhanced effective attraction lies in the increased propensity for solvent depletion at higher salt concentrations. Furthermore, the inter-particle separation-dependent salt density exhibits a similar trend, with Na$^+$ ions consistently showing higher cavity densities than Cl$^-$ ions for hydrophilic (bare and PEG-grafted) Si-NPs, particularly at short separations (see Fig.~S3 in the Supplementary Material). %Finally, as shown in Figs.~S4 and~S5 of the Supplementary Material, the behavior of $U_{\rm eff}$ at elevated ($T,P$) also tracks closely with the solvent depletion propensity within the cavity region.  
\begin{figure*}
    \centering
    \includegraphics[width=0.7\linewidth]{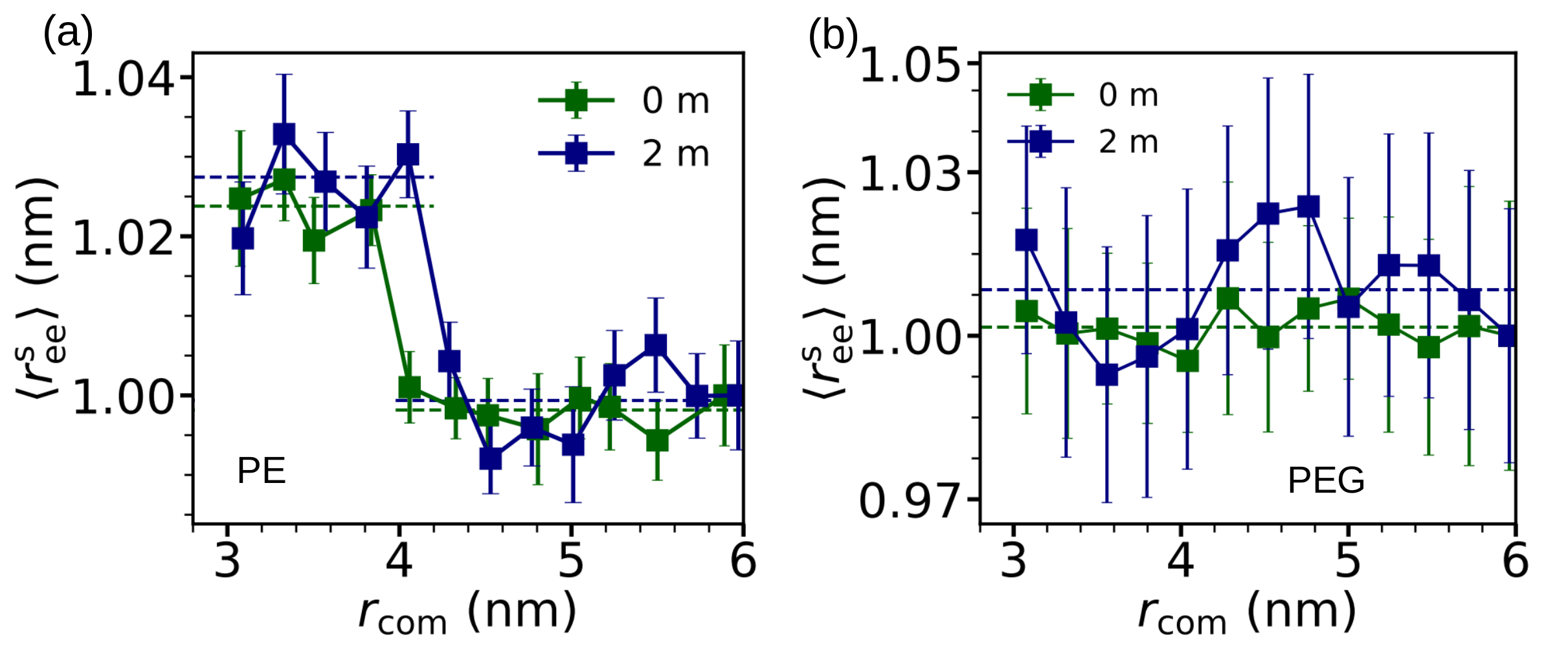}
\caption{Scaled mean end-to-end distance ($\langle r_{\rm ee}^{\rm s} \rangle$) of the grafting PE (a) and PEG (b) polymers as a function of inter-Si-NP center-of-mass separation ($r_{\rm com}$) at temperature $300$ K and $1$ bar pressure. We have reported here the results for $0$ m (pure water) and $2$ m salt concentrations. Error bars represent standard deviations across $5$ independent simulation trajectories. The dashed lines represent the average values of $\langle r_{\rm ee}^{\rm s} \rangle$ calculated over the specified ranges of $r_{\rm com}$. For the PE-grafted case, the short PE chains undergo a transition to relatively elongated conformational states on the onset of the cavitation transition (see Fig.~\ref{fig5}b). This transition gets more pronounced on adding salt to water. The PEG chains, however, show enhanced conformational fluctuations but no significant conformational change on adding salt.}
    \label{fig6}
\end{figure*}
\subsection{Enhanced salt-induced cavity solvent depletion: molecular origin} 
To gain a deeper understanding of the molecular origin of the enhanced solvent depletion propensity from the cavity region upon salt addition for both bare and grafted Si-NPs, we examined in detail the spatial arrangement of water and ions inside the cavity. In Figs.~\ref{fig4}(i) and~\ref{fig4}(ii), we present the scaled (with respect to the corresponding bulk density) average water and ion density profiles along the axis connecting the centres of mass of the two Si-NPs for \(r_{\rm com} = 3.0\)~nm and \(4.0\)~nm, respectively, at $300$ K, $1$ bar and a salt concentration of $2$ m. We find that the ion and water density profiles are strongly influenced by the nature of the grafting polymer. For bare Si-NPs, the Na$^+$ and water densities are enhanced in the vicinity of the Si-NP surface, whereas the Cl$^-$ ions are repelled toward the bulk region (away from the NP surface). For the PEG-grafted Si-NPs, the Na$^+$ density profile displays two pronounced peaks compared to the bare Si-NP case, indicating enhanced accumulation of Na$^+$ ions near the PEG-grafted surfaces for both inter-Si-NP separations. This accumulation can be attributed to the preferential adsorption of Na$^+$ ions on the PEG chains and the Si-NP surface, arising from favorable electrostatic interactions with the electronegative oxygen atoms of PEG and Si-NPs (see Fig.~S4 in the Supplementary Material). The higher Na$^+$ density relative to Cl$^-$ in the cavity, discussed in the previous section (see also Fig.~S3 of the Supplementary Material) can also be attributed to this preferential adsorption of Na$^+$ ions on PEG chains and Si-NP surface. The region near the cavity centre does not exhibit polymer-mediated preference for Na$^+$, leading to lower Na$^+$ density there. The observed enhancement of water and Cl$^-$ density near the cavity center, relative to the Si-NP surface, can be attributed to excluded-volume effects imposed by PEG chains near the Si-NP surface, together with the partial absence of preferential interaction sites for these species. The selective adsorption of Na$^+$ on PEG chains screen the Si-NP surface polymer region from water by disrupting PEG-water hydrogen bonding (Na$^+$ ions compete with water for interactions with PEG). This process leads to hydration repulsion: the PEG polymer region becomes less hydrated, the inter-particle cavity loses water and the effective inter-particle attraction is consequently enhanced --- analogous to cavitation-induced attraction between nanoscale hydrophobic surfaces. 

In contrast, for PE-grafted Si-NPs, favorable salt-polymer  interaction sites are absent on PE chains. As a result, salt ions interact more strongly with water molecules than with the PE chains. This effectively removes water molecules from the vicinity of the PE-grafted Si-NPs, consistent with the well-known salting-out effect~\cite{hey2005salting, sadeghi2012salting, endo2012salting, kalra2001salting, bruce2021contact, chudoba2018tuning, kesselman2002atr}. The salting-out process reduces the solubility of the hydrophobic PE segments, enhances solvent expulsion from the inter-particle cavity, and thus, amplifies the effective inter-particle interaction. The weak salt-polymer interaction also results in a significantly lower salt density in the cavity region at both inter-Si-NP separations (Fig.~\ref{fig4}(i-ii)b). Overall, although the molecular mechanisms differ --- hydration repulsion due to preferential ion adsorption in the PEG case versus salting-out in the PE case --- salt addition promotes a common outcome: enhanced solvent depletion from the inter-particle cavity and strengthening the effective interaction between the grafted Si-NPs. 
\subsection{Manifestations of partial cavitation transition in the conformational behavior of grafting polymers}
It is well established that the conformational properties of polymers are highly sensitive  to both solvent characteristics and spatial confinement~\cite{williams1981polymer,kraus2000confinement}. The results reported above (Sections~IVC and~IVD) suggest that cavity solvent properties are  significantly altered when the inter-particle separation is reduced, and this alteration may lead to changes in the conformational behavior of the grafted polymers. Therefore, it is  natural to probe whether signatures of solvent-induced changes in the effective Si-NP interaction are also reflected in the conformational behavior of grafted polymers at different salt concentrations. To quantify the conformational behavior of grafted polymers, we computed the scaled average end-to-end distance of the grafted PE and PEG chains ($\langle r_{\rm ee}^{\rm s} \rangle$), defined as $\langle r_{\rm ee}^{\rm s} \rangle = \langle r_{\rm ee} \rangle / \langle r_{\rm ee}^{\rm o} \rangle$, as a function of the inter-Si-NP separation $r_{\rm com}$ at $0$ m and $2$ m salt concentration. Here, $\langle r_{\rm ee} \rangle$ is the average end-to-end distance at a given $r_{\rm com}$, and $\langle r_{\rm ee}^{\rm o} \rangle$ corresponds to the value at which the NPs are maximally separated ($r_{\rm com} \sim 6$~nm) in the cubic simulation box with an average box length of approximately $12$~nm. 

The conformational behavior of the grafted polymers, characterized by $\langle r_{\rm ee}^{\rm s} \rangle$,  exhibits distinct trends for PE and PEG-grafted Si-NPs (see Fig.~\ref{fig6}). For PE-grafted Si-NPs, a sharp transition in $\langle r_{\rm ee}^{\rm s} \rangle$ is observed near $r_{\rm com} \approx 4$~ nm, which coincides with the onset inter-particle separation where the cavitation transition is observed for both salt concentrations (Fig.~\ref{fig5}b) --- consistent with our previous study in salt-free solvent~\cite{singh2024computational} (the cavitation transition leads to the stretching of PE chains due to the significantly reduced solvent density which allows the chains to extend more freely). This conformational transition becomes marginally more pronounced with increasing salt concentration, correlating with the salt-induced enhancement of the cavitation transition (Fig.~\ref{fig5}b) and, in turn, the strengthening of effective inter-particle attraction (Fig.~\ref{fig:1}b). Thus, the conformational behavior of the grafted PE chains provides additional evidence for the underlying origin of the salt-induced enhancement of effective (PE-grafted) Si-NP interactions. In contrast, PEG-grafted Si-NPs show no abrupt conformational change. Instead, they exhibit increased conformational fluctuations and a slight overall increase in end-to-end distance upon salt addition. This behavior suggests that a more detailed analysis of charge screening effects and polymer solubility is required to fully elucidate the underlying conformational response to solvent conditions and confinement. 
\section{Conclusions}
In this study, we employed all-atom MD simulations to systematically investigate how the solvent-mediated effective interaction, $U_{\rm eff}$, between two Si-NPs in an aqueous solution is influenced by S-NP’s surface functionalization under varying salt (NaCl) concentrations and thermodynamic conditions. Specifically, we examined the interplay between surface functionalization, solvent ionic strength, and thermodynamic conditions on the pair effective interaction. Our results reveal that salt affects the interactions of different surface-functionalized Si-NPs in distinct ways. For bare (ungrafted) Si-NPs, $U_{\rm eff}$ shows negligible changes with increasing salt concentration. In contrast, PE-grafted Si-NPs exhibit a pronounced increase in inter-particle attraction even at a moderate ($2$ m) salt concentration, which we attribute to salting-out-induced enhancement of cavitation in the inter-particle cavity. PEG-grafted systems, on the other hand, display a more gradual deepening of $U_{\rm eff}$ with increasing salt concentration. This behavior arises from ion-induced screening of the hydrophilic PEG chains through selective Na$^+$ adsorption, which weakens the water–PEG hydrogen bonds and reduces the water density in the cavity compared to salt-free water, particularly at near-contact inter-particle separations.

To asses the transferability of these salt-induced effects on effective interactions, we further examined the effects of salt on $U_{\rm eff}$ under different ($T, P$) conditions. We found that the salt-effects remain largely insensitive to variations in $T$ and $P$ (in the range reported in this work). To gain further insight, we analyzed the conformational response of the grafted polymers to cavity solvent depletion. For PE-grafted systems, we observed a sharp increase in chain extension near the onset of cavitation transition inter-particle separation ($r_{\rm com} \approx 4$ nm), which is amplified at higher salt concentrations. This coincides with the steep drop in $U_{\rm eff}$ at $r_{\rm com} \approx 4$~nm and the enhanced inter-Si-NP attraction upon salt addition. In contrast, PEG chains exhibit enhanced conformational fluctuations and only a marginal increase in chain length with rising salt concentration.  

To summarize, our results underscore the complex and cooperative mechanisms by which solvent composition, and surface chemistry govern $U_{\rm eff}$ between nanoscale particles dispersed in solution. Such understanding, and the ability to predictably control $U_{\rm eff}$, contributes to broader efforts in inverse materials design, enabling the programmable self-assembly of nanoscale building blocks through tailored surface functionalization and solvent engineering.

\section{Acknowledgements}
 R.S.S. acknowledges financial support from DST-SERB (Grant No. CRG/2023/002975). Y.S. acknowledges financial support from IISER Tirupati. The computations were performed at the IISER Tirupati computing facility and at PARAM Brahma at IISER Pune.

\section*{References}
\bibliography{achemso-demo}

\section*{Reference}
\begin{thebibliography}{1}
\bibitem{SI_ref_yu2024interactions}
~M.~Yu, X.~Kang, and L.~Qian,
``Interactions between monovalent cations and polyethylene glycol: A study at micro level,''
\textit{Colloids and Surfaces A: Physicochemical and Engineering Aspects}
\textbf{680}, 132731 (2024).
\end{thebibliography}

%%%%%%%%%% Merge with supplemental materials %%%%%%%%%%
\pagebreak
\widetext
\begin{center}
\textbf{\large Supplementary Material}
\end{center}
%%%%%%%%%% Merge with supplemental materials %%%%%%%%%%
%%%%%%%%%% Prefix a "S" to all equations, figures, tables and reset the counter %%%%%%%%%%
\setcounter{equation}{0}
\setcounter{figure}{0}
\setcounter{table}{0}
\setcounter{page}{1}
\setcounter{section}{0}
\makeatletter
\renewcommand{\theequation}{S\arabic{equation}}
\renewcommand{\thefigure}{S\arabic{figure}}
\renewcommand{\bibnumfmt}[1]{[S#1]}
\renewcommand{\citenumfont}[1]{S#1}
%%%%%%%%%% Prefix a "S" to all equations, figures, tables and reset the counter %%%%%%%%%%

\begin{figure}[h]
    \centering
    \includegraphics[width=0.4\linewidth]{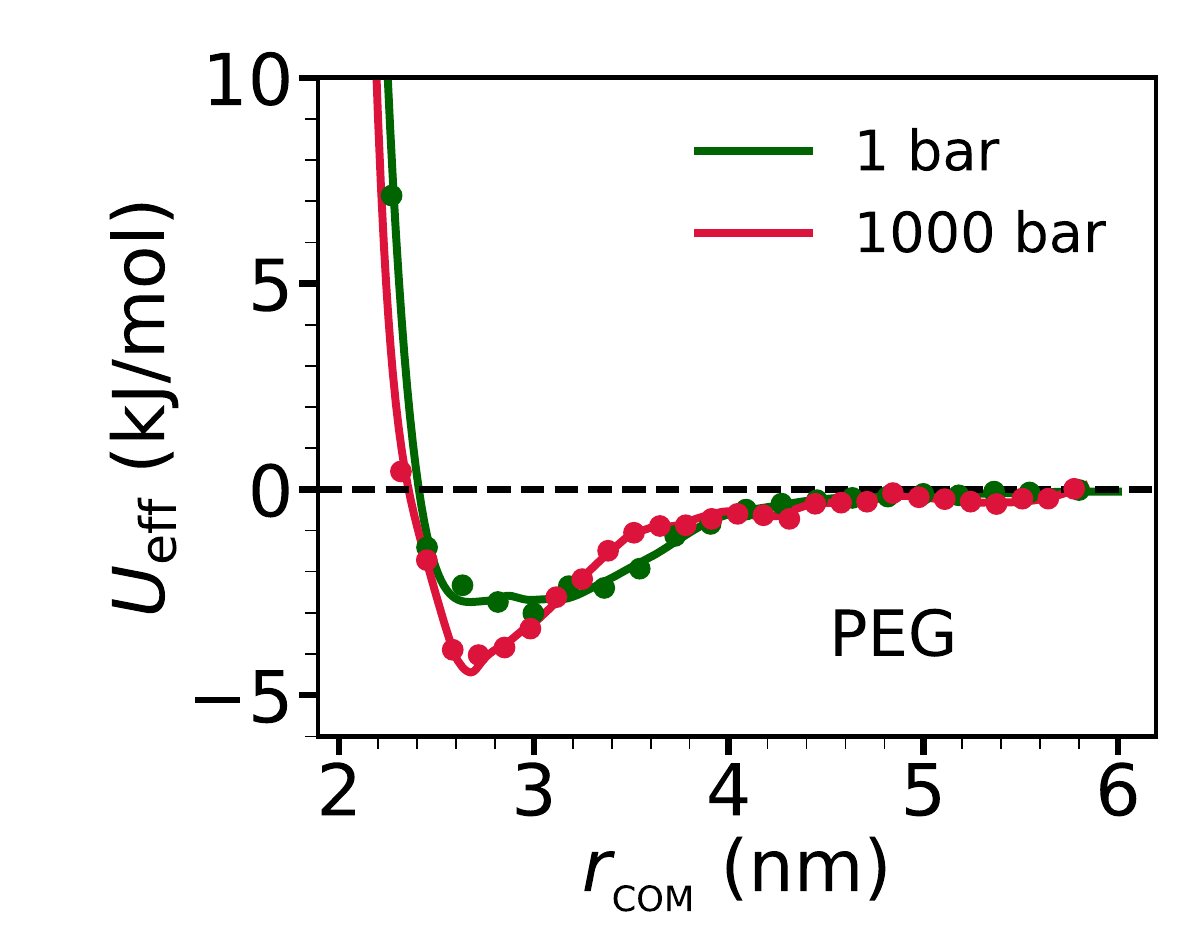}
    \caption{The effects of pressure change on $U_{\rm eff}$ between two PEG-grafted Si-NPs at $300$ K in a salt-free solvent is presented. Similar to the bare Si-NP case, the $U_{\rm eff}$ between PEG-grafted S-NPs exhibits only a weak dependence on pressure change.}
    \label{fig:SI-1}
\end{figure}

\begin{figure}[h]
    \centering
    \includegraphics[width=0.4\linewidth]{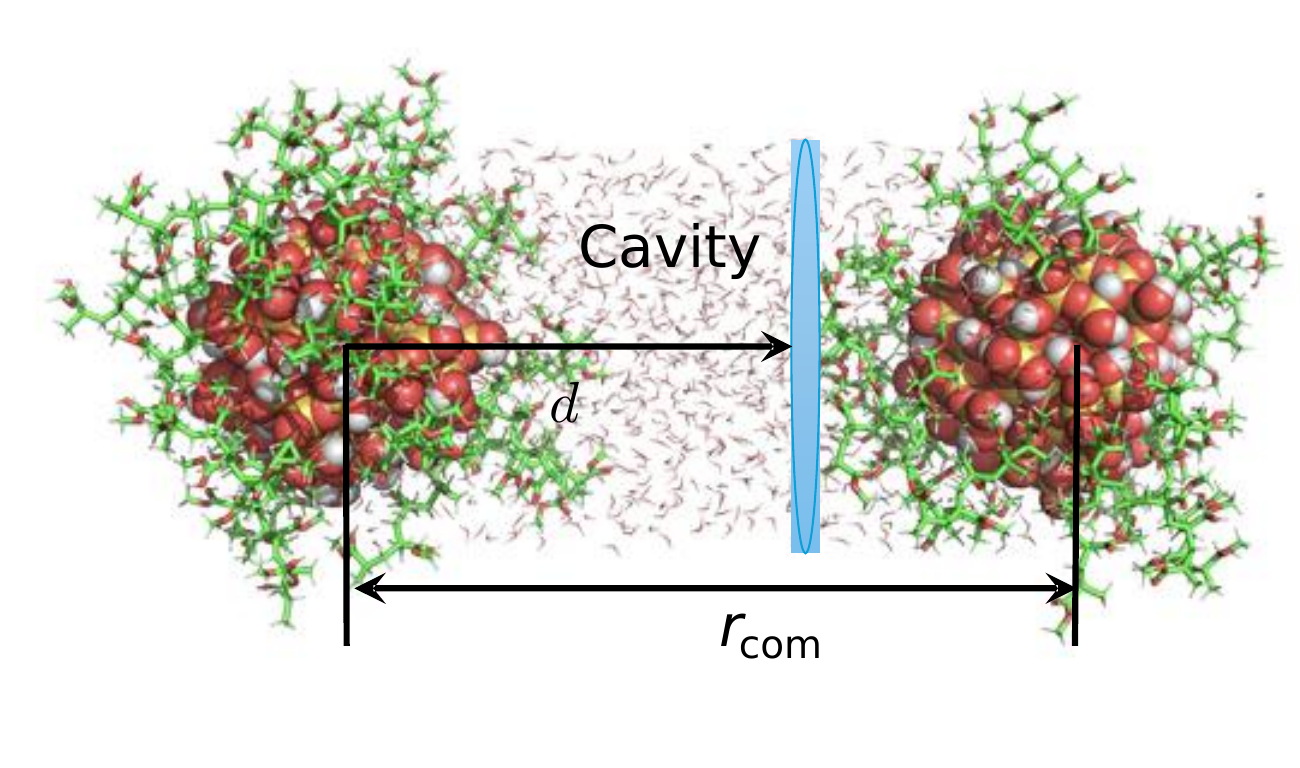}
    \caption{Polymer-grafted Si--NPs are shown with solvent molecules occupying the interparticle cavity region. 
The distance between the centers of mass of the two nanoparticles is denoted as $r_{\rm com}$.  The vector $d$ is defined as the distance from the center of mass of one nanoparticle pointing towards the center of mass of the other.}
    \label{fig:SI-2}
\end{figure}
\begin{figure}[H]
    \centering
    \includegraphics[width=0.6\linewidth]{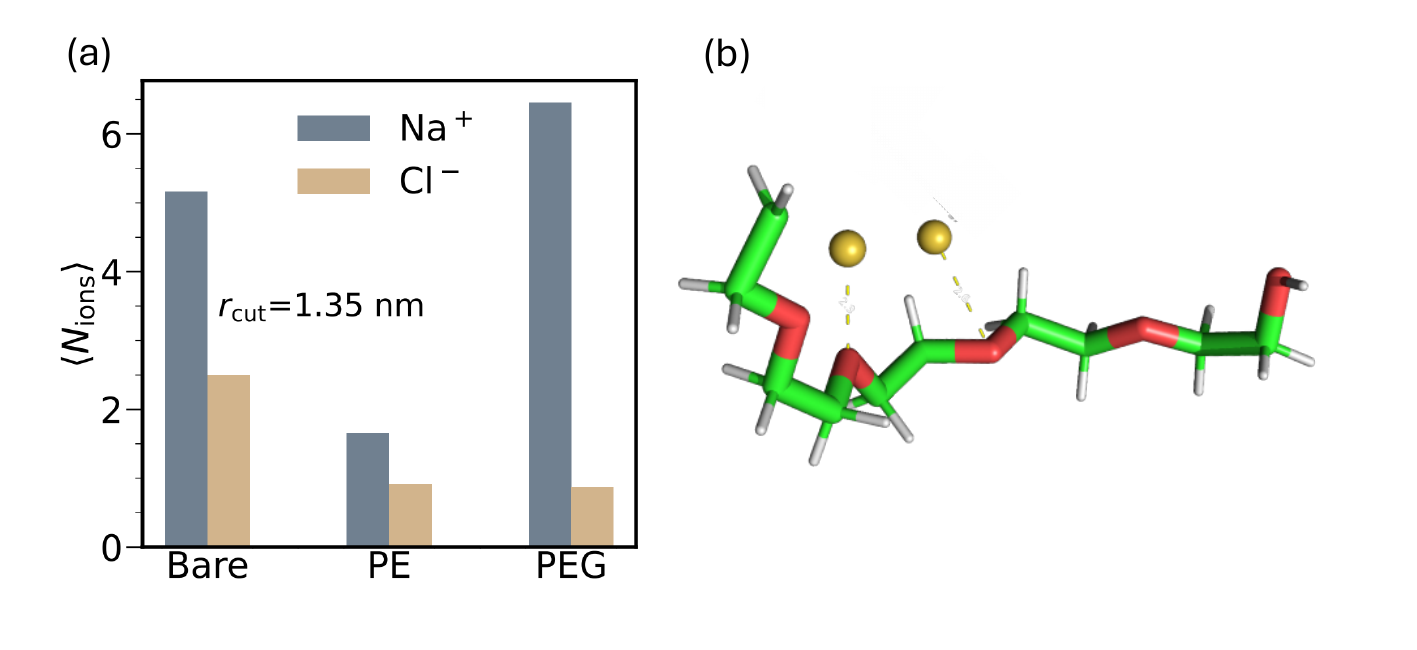}
\caption{Ion structuring in the cavity formed by two PE-grafted Si-NPs. (a) To further understand the propensity of salt-polymer interactions, we show the number of ions inside the spherical volume defined by radial cutoff distances of $r_{\rm cut} = 1.35$ nm from the Si-NP center at $T = 300$~K, $P = 1$~bar, and 2~m salt concentration. At $r_{\rm cut} = 1.35$~nm, the number of Na$^+$ ions exceeds that of Cl$^-$, with the PEG-grafted system showing the most significant difference. (b) We show a representation of Na$^+$ ions in the vicinity of a single PEG chain (red: ether oxygen, green: carbon, white: hydrogen). For clarity, Si-NP, water, and Cl$^-$ have been omitted from the image.}
    \label{fig:SI-3}
\end{figure}
\begin{figure}[h]
    \centering
    \includegraphics[width=0.8\linewidth]{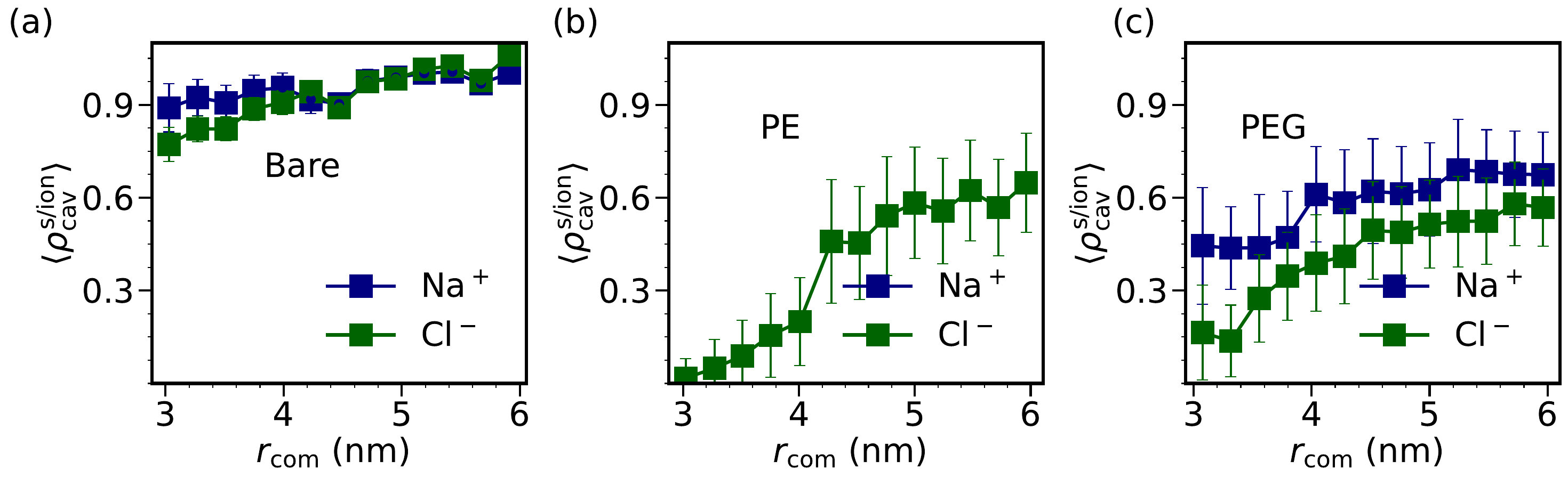}
    \caption{Scaled ion number density in the cavity ($\rho_{\rm cav}^{\rm s/ion}$) vs $r_{\rm com}$ for bare (a), PE-grafted (b), and PEG-grafted (c) Si-NPs at 1 bar and 300 K in the presence of 2 m salt concentration. The $\langle \rho_{\rm cav}^{\rm s/ion} \rangle$ of Cl$^-$ decreases with decreasing $r_{\rm com}$ across all systems. In the PE-grafted system, both Na$^+$ and Cl$^-$ exhibit a decrease in $\langle \rho_{\rm cav}^{\rm s/ion} \rangle$ with decreasing $r_{\rm com}$, showing a sharp drop around $r_{\rm com} = 4$~nm, and eventually escaping the cavity near $r_{\rm com} = 3.0$~nm. In contrast, for hydrophilic Si-NPs (i.e., bare and PEG-grafted), Na$^+$ ions relatively tend to remain near the surface, indicating a relatively stronger propensity for interaction with the surface compared to Cl$^-$ ions. These observations are aligned with recent experimental~\cite{SI_ref_yu2024interactions} findings, which showed that smaller monovalent cations such as Li$^+$ and Na$^+$ exhibit stronger interactions with PEG chains.}
    \label{fig:SI-4}
\end{figure}

\cleardoublepage
\end{document}